\documentclass[aps,prb,twocolumn,longbibliography,showpacs,superscriptaddress]{revtex4-1}

\usepackage[T1]{fontenc}
\usepackage{latexsym}
\usepackage{graphicx} 
\usepackage{epstopdf}
\usepackage{amsmath}  
\usepackage{amssymb}
\usepackage{cases}
\usepackage{lipsum}
\usepackage{float}
\usepackage{hyperref}

\begin{document}

\title{Spin-triplet paired phases inside ferromagnet induced by Hund's rule coupling and electronic correlations: Application to $\mathbf{UGe_2}$}
\author{E. K\k{a}dzielawa-Major}
 \email{ewa.kadzielawa@doctoral.uj.edu.pl}
\affiliation{Marian Smoluchowski Institute of Physics, Jagiellonian University, ul. {\L}ojasiewicza 11, 30-348 Krak{\'o}w, Poland }
 \author{M. Fidrysiak}
 \email{maciej.fidrysiak@uj.edu.pl}
\affiliation{Marian Smoluchowski Institute of Physics, Jagiellonian University, ul. {\L}ojasiewicza 11, 30-348 Krak{\'o}w, Poland }
 \author{P. Kubiczek}
 \email{patryk.kubiczek@physik.uni-hamburg.de}
\affiliation{I. Institut f{\"u}r Theoretische Physik, Universit{\"a}t Hamburg, Jungiusstra{\ss}e 9, D-20355 Hamburg, Germany}
 \author{J. Spa{\l}ek}%
 \email{jozef.spalek@uj.edu.pl}
\affiliation{Marian Smoluchowski Institute of Physics, Jagiellonian University, ul. {\L}ojasiewicza 11, 30-348 Krak{\'o}w, Poland }

\date{\today}

\begin{abstract}
We discuss a mechanism of real-space spin-triplet pairing, alternative to that due to quantum paramagnon excitations, and demonstrate its applicability to $\mathrm{UGe_2}$. Both the Hund's rule ferromagnetic exchange and inter-electronic correlations contribute to the same extent to the equal-spin pairing, particularly in the regime in which the weak-coupling solution does not provide any. The theoretical results, obtained within the orbitally-degenerate Anderson lattice model, match excellently the observed phase diagram for  $\mathrm{UGe_2}$ with the coexistent ferromagnetic (FM1) and superconducting ($A_1$-type) phase. Additionally, weak $A_2$- and $A$-type paired phases appear in very narrow regions near the metamaganetic (FM2 $\rightarrow$ FM1) and FM1 $\rightarrow$ paramagnetic first-order phase-transition borders, respectively. The values of magnetic moments in the FM2 and FM1 states are also reproduced correctly in a semiquantitative manner. The Hund's metal regime is also singled out as appearing near FM1-FM2 boundary.
\end{abstract}

\maketitle

\section{Introduction}
\label{sec:introduction}

The discovery of superconductivity (SC) in uranium compounds $\mathrm{UGe_2}$,\cite{SaxenaNature2000,TateiwaJPhysCondensMatter2001,PfleidererPhysRevLett2002,HuxleyPhysRevB2001} $\mathrm{URhGe}$,\cite{AokiNature2001} $\mathrm{UCoGe}$,\cite{HuyPhysRevLett2007} and $\mathrm{UIr}$ \cite{KobayashiPhysicaB2006} that appears inside the ferromagnetic (FM) phase, but close to magnetic instabilities, has reinvoked the principal question concerning the mechanism of the spin-triplet pairing.  The latter is particularly intriguing, since the spin-triplet SC \cite{PfleidererRevModPhys2009,AokiJPhysSocJapan2012,MackenzieRevModPhys2003,HuxleyPhysicaC2015} occurs relatively seldom in the correlated systems as compared to its spin-singlet analogue. More importantly, the circumstance that the paired state in both $\mathrm{UGe_2}$ and $\mathrm{UIr}$ is absent on the paramagnetic (PM) side of the FM1$\rightarrow$PM discontinuous transition, suggests a specific mechanism providing on the same footing both the magnetic and SC orderings. Moreover, SC is well established in one particular (FM1) magnetic phase, but not in FM2 or PM phases, where the magnetic moment is either almost saturated or vanishes, respectively. These circumstances pose a stringent test on any pairing mechanism which should be tightly connected to the onset/disappearance of ferromagnetism.

The spin-triplet SC mediated by the quantum spin fluctuations has been invoked \cite{ColemanNature2000,SandemanPhysRevLett2003} and tested for $\mathrm{UCoGe}$ \cite{HattoriPhysRevLett108,TadaJPhysConfSeries2013,WuNatCommun2017} that represents the systems with very low magnetic moments\cite{SatoAIPConfProc2011,HattoriPhysRevLett108} ($m \sim 0.039 \, \mu_B / \mathrm{U}$) and thus is particularly amenable to the fluctuations in both the weakly-ordered FM and PM regimes. From this perspective, $\mathrm{UGe_2}$ possesses a large magnetic moment in FM1 phase ($m \sim 1 \, \mu_B / \mathrm{U}$), and in the low-pressure FM2 phase it is even larger ($m \sim 1.5 \, \mu_B / \mathrm{U}$).\cite{PfleidererPhysRevLett2002} In such a situation, a natural idea arises that in this case local correlation effects should become much more pronounced in $\mathrm{UGe_2}$, particularly because the dominant SC phase appears in between two metamagnetic transitions, one of which ($\mathrm{FM2}\rightarrow\mathrm{FM1}$) can be associated with the transition from almost localized FM2 phase of $5f$ electrons. Closely related to this is the question of real-space spin-triplet pairing applicability, considered before as relevant to the orbitally-degenerate correlated narrow-band systems,\cite{SpalekPhysRevB2001,KlejnbergJPhysCondensMatter1999,SpalekSpringer2002,SpalekJPhysCondensMatter2013,ZegrodnikNewJPhys2013,ZegrodnikNewJPhys2014,HanPhysRevB2004,HoshinoPhysRevLett2015,DaiPhysRevLett2008} which in turn is analogous to the spin-singlet pairing proposed for the high-temperature \cite{AndersonJPhysCondensMatter2004,EdeggerAdvPhys2007,OgataRepProgPhys2008,JedrakPhysRevB2011,KaczmarczykNewJPhys2014,SpalekPhysRevB2017} and heavy-fermion \cite{BodensiekPhysRevLett2013,HowczakPhysStatSolidiB2013,WuPhysRevX2015,WysokinskiPhysRevB2016} superconductors. Essentially, we explore the regime of large and weakly fluctuating moments. The relevance of this idea is supported by the recent experimental evidence that the ratio of spontaneous moment $m$ to its fluctuating counterpart $m_0$ is $\sim 1$, whereas for $\mathrm{UCoGe}$ $m_0 \gg m$, so the two systems are located on the opposite sides of the Rhodes-Wohlfarth plot.\cite{TateiwaPhysRevB2017}

Explicitly, we put forward the idea of the correlation-induced pairing and test it for the case of $\mathrm{UGe_2}$. To implement that program we generalize our approach, applied earlier \cite{WysokinskiPhysRevB2014,WysokinskiPhysRevB2015,AbramJMagnMagnMat2016} to explain the magnetic properties of $\mathrm{UGe_2}$, and incorporate this specific type of the coexistent SC into that picture. Specifically, we extend the spin-triplet pairing concepts, originally introduced for the case of multi-orbital narrow-band systems,\cite{SpalekPhysRevB2001,KlejnbergJPhysCondensMatter1999,SpalekSpringer2002,SpalekJPhysCondensMatter2013,ZegrodnikNewJPhys2013,ZegrodnikNewJPhys2014,HanPhysRevB2004} by including the Hund's rule coupling combined with intraatomic correlations within the orbitally-degenerate Anderson lattice model (ALM), and treat it within the statistically consistent version of the renormalized mean-field theory (SGA).\cite{WysokinskiPhysRevB2014,WysokinskiPhysRevB2015,AbramJMagnMagnMat2016} In this manner, we demonstrate, in quantitative terms, the applicability of the concept of even-parity, spin-triplet pairing to $\mathrm{UGe_2}$. Furthermore, we provide also a detailed analysis of the two very narrow border regions FM2-FM1 and FM1-PM, in which a weak $A_2$-type SC transforms to $A_1$ and from $A_1$ to practically marginal $A$ phase, respectively, before SC disappears altogether (the notation of the SC phases is analogous\cite{WheatleyRevModPhys1975} to that used for superfluid $^3$He).

The present mechanism may be regarded as complementary to the reciprocal-space pairing by long-wavelength quantum spin fluctuations which was very successful in explaining the properties of the superfluid $\mathrm{{}^3He}$.\cite{AndersonPhysRevLett1973,BrinkmanPhysRevA1974} The latter mechanism was also applied to ferromagnets with magnetic moment fluctuations, both on the weakly-FM and PM sides.\cite{FayPhysRevB1980,MonthouxPhysRevB1999} Specifically, the role of their longitudinal component was emphasized. However, all those considerations have been limited to a single-band situation and therefore, SC is unavoidably of the $p$-wave character. The multiband structure, considered here, allows for even-parity SC state which can take the form of $s$-wave.

\section{Model and method}
\label{sec:model_and_method}

We start with doubly degenerate $f$ states and assume two-dimensional structure of the compound.\cite{ShickPhysRevLett2001,ShickPhysRevB2004} Within our model, the total number of electrons per formula unit $n^\mathrm{tot} \equiv n^f + n^c$, with $n^f$ and $n^c$ being the $f$ and conduction ($c$) electron occupancies, must be exceeding that on $5f$ level for $\mathrm{U^{3+}}$ ion,\cite{PfleidererRevModPhys2009,TrocPhysRevB2012,SamselCzekalaIntermetallics2011} i.e., $n > 3$. The best comparison with experiment is here achieved for $n^\mathrm{tot} \simeq 3.25$. This presumption brings into mind the idea of an orbitally selective delocalization of one out of the three $5f$ electrons under pressure (see below).

Explicitly, we employ a four-orbital ALM defined by the Hamiltonian (with the chemical potential term $-\mu \hat{N}_e$ included) 

\begin{align}
  \label{eq:hamiltonian}
  \mathcal{H} - &\mu \hat{N}_e = \sum_{i j l \sigma} t_{ij} \hat{c}^{(l)\dagger}_{i\sigma} \hat{c}^{(l)}_{j\sigma} + V \sum_{i l \sigma} \left( \hat{f}^{(l)\dagger}_{i\sigma} \hat{c}^{(l)}_{i\sigma} + \mathrm{H.c.}\right) + \nonumber\\ & + \epsilon^f \sum_{i l} \hat{n}^{f(l)}_i + U \sum_{il} \hat{n}^{f(l)}_{i\uparrow} \hat{n}^{f(l)}_{i\downarrow} + U' \sum_{i} \hat{n}^{f(1)}_i \hat{n}^{f(2)}_i - \nonumber \\ & - 2 J \sum_i \left( \mathbf{\hat{S}}_i^{f(1)} \cdot \mathbf{\hat{S}}_i^{f(2)} + \frac{1}{4} \hat{n}_i^{f(1)} \hat{n}_i^{f(2)} \right) - \mu \hat{N}_e,
\end{align}

\noindent
involving two $f$-orbitals (with creation operators $\hat{f}^{(l)\dagger}_{i\sigma}$ with $l = 1, 2$ at lattice site $i$ and spin $\sigma = \uparrow, \downarrow$), hybridized with two species of conduction electrons created by $\hat{c}_{i\sigma}^{(l)\dagger}$ (minimally two $c$ bands are needed, as otherwise one of the $f$-orbitals decouples and does not participate in the resultant quasiparticle states \cite{SpalekSolStatCommun1985}). Out of general hopping matrix $t_{ij}$ we retain nearest- and next-nearest neighbor hoppings ($t$, $t'$) and assume local character of $f$-$c$ hybridization $V$. Correlations in the $f$-electron sector are governed by intra-orbital $f$-$f$ repulsion $U$, inter-orbital repulsion $U'$, and Hund's coupling $J$. Here $\hat{n}^{f(l)}_i$, $\mathbf{\hat{S}}^{f(l)}_i$ denote the $f$-electron number and spin operators on site $i$ and for orbital $l$, whereas $\hat{N}_e$ is the total particle number.  Hereafter, we restrict ourselves to the case of $U' = U - 2J$, $U / |t| = 3.5$, and $t' / |t| = 0.25$. The values of parameters have been selected to reproduce correctly the observed values of magnetic moments, the magnetic critical points,\cite{WysokinskiPhysRevB2015} and the maximal value of SC transition temperature $T_\mathrm{SC} \lesssim 1\,\mathrm{K}$, all at the same time. Also, we neglect the interorbital pair-hopping term $\sim J$ as it contributes only to the spin-singlet pairing channel.

The SGA approach is based on optimization of the ground state energy within the class of wave functions with partially projected-out double $f$-orbital occupancies, and can be formulated in terms of effective one-body Hamiltonian 

\begin{align}
	\mathcal{H}_\mathrm{eff}  = \sum_{\mathbf{k}, \sigma}^{} \Psi_{\mathbf{k} \sigma}^\dagger \left( \begin{array}{cccc}
		\epsilon_{ \mathbf{k} } & 0 &  q_\sigma V&  0\\
		0 & - \epsilon_{ \mathbf{k} } & 0 & - q_\sigma V\\
		q_\sigma V & 0 & \epsilon_{ \sigma }^{f} & \Delta_{ \sigma \sigma }^{ff} \\
		0 & - q_\sigma V & \Delta_{ \sigma \sigma }^{ff} & -\epsilon_{ \sigma }^{f} \\
	\end{array}
	\right) \Psi_{\mathbf{k} \sigma}
  + E_0,
  \label{eq:heff}
\end{align}

\noindent
derived from the model of Eq.~\eqref{eq:hamiltonian} (cf. Appendix~\ref{appendix:SGA}). In Eq.~\eqref{eq:heff} $\Psi_{\mathbf{k}\sigma }^\dagger \equiv \left( \hat{c}^{(1)\dagger}_{ \mathbf{k} \sigma }, \hat{c}^{(2)}_{ -\mathbf{k} \sigma }, \hat{f}^{(1)\dagger}_{ \mathbf{k} \sigma }, \hat{f}^{(2)}_{ -\mathbf{k} \sigma } \right)$, $\epsilon_\mathbf{k}$ denotes bare $c$-electron dispersion relation, $\epsilon_\sigma^f$ is an effective $f$-level, $\Delta_{\sigma\sigma}^{ff} \equiv \mathcal{V}_\sigma \langle \hat{f}_{i\sigma}^{(1)} \hat{f}_{i\sigma}^{(2)} \rangle$ is the $f$-$f$ equal-spin SC gap parameter, $\mathcal{V}_\mathrm{\sigma} \equiv  - U'  g_{1\sigma} + (J - U')  g_{2\sigma}$ denotes effective pairing coupling, and $E_0$ is a constant. The renormalization factors $q_\sigma$, $g_{1\sigma}$, and $g_{2\sigma}$ account for the correlation effects and originate from projection of the trial wave functions (see Appendix~\ref{appendix:SGA} for explicit expressions).

The basic quantity determined from  the diagonalization of $\mathcal{H}_\mathrm{eff}$ (cf. Appendix~\ref{appendix:SGA}) is the quasiparticle gap $\Delta_\mathbf{k}$. For wave vectors lying on the Fermi surface of the normal-state, one obtains

\begin{align}
\Delta_\mathbf{k}^2 = \frac{\epsilon_\mathbf{k}^2}{(\epsilon_\mathbf{k} + \epsilon_\sigma^f)^2} \times (\Delta_{\sigma\sigma}^{ff})^2 + o[(\Delta_{\sigma\sigma}^{ff})^2],
\end{align}

\noindent
so $\Delta_\mathbf{k}$ is expressed in terms $\Delta_{\sigma\sigma}^{ff}$ and a weakly $\mathbf{k}$-dependent factor. Therefore, in the remaining discussion we use the latter gap, underlying in this manner the dominant role of the $f$-$f$ pairing.

The quantity particularly relevant to the present discussion, is the equal-spin coupling constant $\mathcal{V}_\sigma$. If positive, this term favors equal-spin triplet SC. We also define the Hartree-Fock (HF/BCS) coupling constant $\mathcal{V}_\mathrm{HF} = J - U'$, independently of the spin direction. In the latter approximation the interatomic interaction is attractive when $J - U' = 3J - U > 0$ (this condition defines the BCS limit). One of the principal signatures of correlation importance is that \textit{pairing persists even when the coupling $\mathcal{V}_\mathrm{HF}$ becomes repulsive} ($\mathcal{V}_\mathrm{HF} < 0$), as shown below. The conditions $\mathcal{V}_\mathrm{HF} < 0$ and $\mathcal{V}_\sigma > 0$ define the regime of correlation-driven SC.

\begin{figure}
   \centering
\includegraphics[width = \columnwidth]{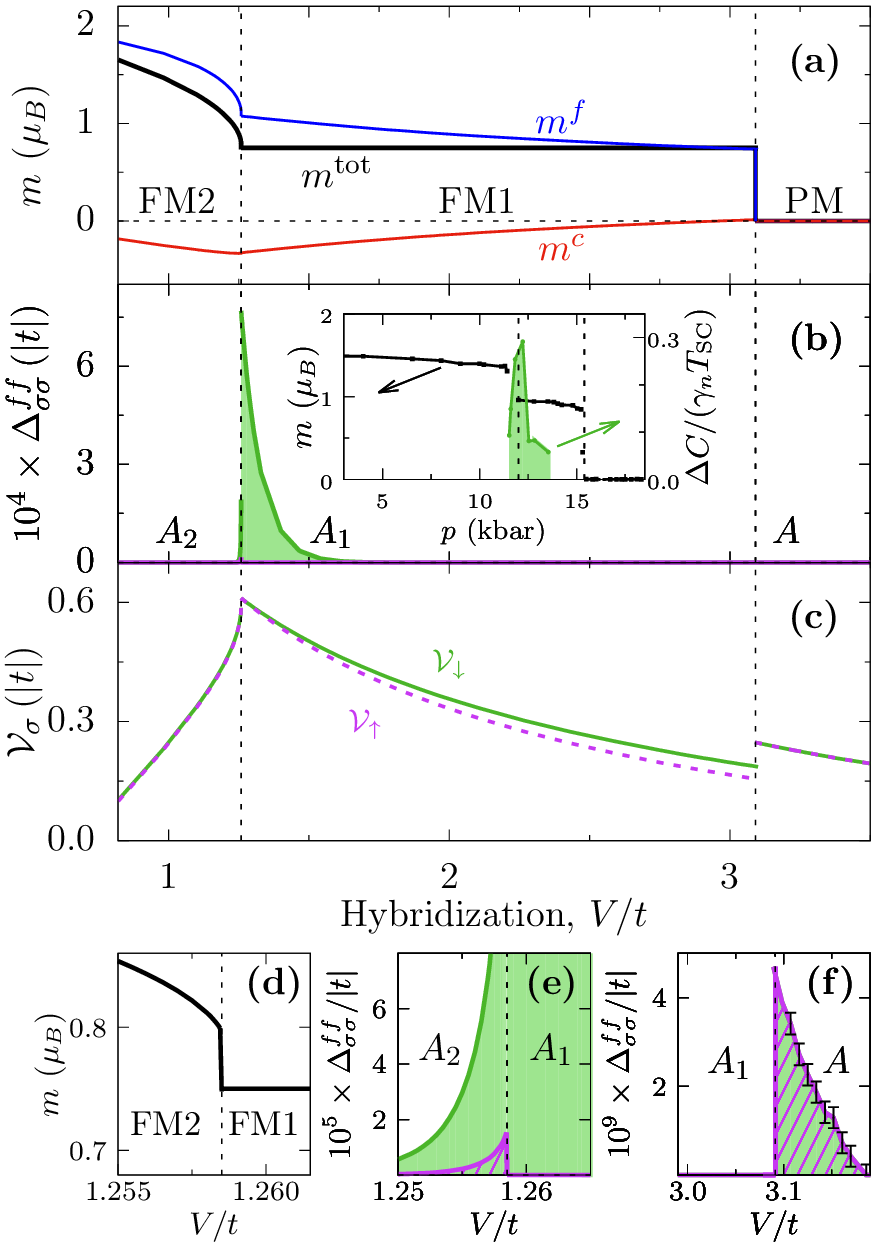} %
\caption{Calculated zero-temperature phase diagram of $\mathrm{UGe_2}$ for  Hund's coupling $J / |t|= 1.1$ versus $f$-$c$ hybridization $V$. The remaining parameters read: $t' /|t| = 0.25$, $U / |t|= 3.5$, $\epsilon^f / |t|= -4$, and $n^\mathrm{tot} = 3.25$. (a) Total magnetic moment $m^\mathrm{tot}$ per formula unit (black solid line), and the corresponding $f$ and $c$ electron magnetization $m^f$ and $m^c$ (blue and red lines, respectively). $m^c$ represents a residual Kondo compensating cloud. (b) Triplet $f$-$f$ SC gap component $\Delta^{ff}_{\uparrow\uparrow}$ (purple shading) and $\Delta^{ff}_{\downarrow\downarrow}$ (green shading). Three distinct SC phases  $A_2$, $A_1$, and $A$ are marked. The $A$-phase gaps ($\sim 10^{-9}|t|$) are not visible in panel (b). Inset shows experimental  magnetization for $\mathrm{UGe_2}$ \cite{PfleidererPhysRevLett2002} and the specific-heat jump at the SC transition temperature $T_\mathrm{SC}$ (normalized by $T_\mathrm{SC}$ and the linear specific-heat coefficient $\gamma_n$).\cite{TateiwaPhysRevB2004} (c) Effective coupling constant $\mathcal{V}_\sigma$ for spin-up (purple) and spin-down (green) triplet pairing. Note that value of coupling is the largest near the $A_2 \rightarrow A_1$ transition. (d) Total magnetic moment near the FM2$\rightarrow$FM1 metamagnetic transition. (e)-(f) SC gap components near the FM2$\rightarrow$FM1 and FM1$\rightarrow$PM transition points, respectively.}%
  \label{fig:phase_diagram}
\end{figure}

 \begin{figure}
   \centering
   \parbox{1\columnwidth}{\includegraphics[width = 1\columnwidth]{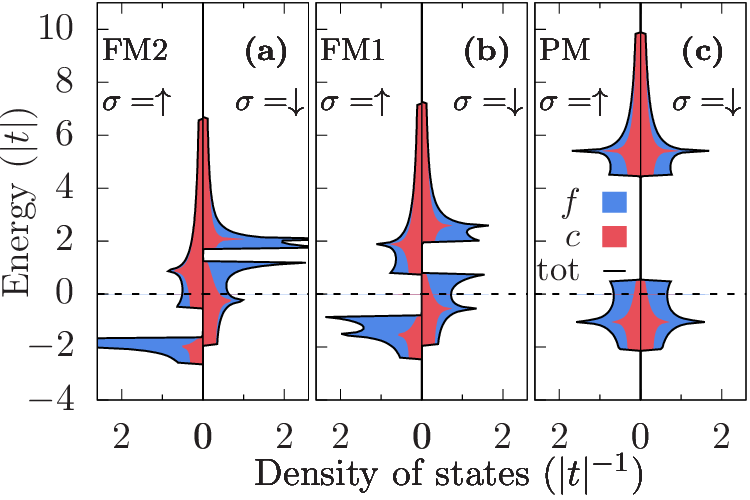}} %
   \caption{Spin- and orbital-resolved density of states for $J/|t| = 1.1$ in the (a) FM2 ($V/t = 1.1$), (b) FM1 ($V/t = 1.625$), and (c) PM ($V/t = 3.25$) phases. Orbital contributions are marked in blue and red, whereas the total density of states is plotted by black solid line. Dirac-delta functions have been smeared out by $\epsilon = 10^{-3}|t|$ for numerical purposes.}%
  \label{fig:dos}
\end{figure}

\section{Results}
\label{sec:results}

The complete phase diagram encompassing both FM and SC states, for selection of Hund's coupling $J/|t| = 1.1$, is shown in Fig.~\ref{fig:phase_diagram} (see Appendices~\ref{appendix:numerical_procedure}-\ref{appendix:phase_diagram} for technical aspects of the analysis). In panel (a) we exhibit the system evolution from the large-moment FM2 phase, through FM1 state with a magnetization plateau at $\sim 0.8 \mu_B$ (as compared to $\sim 1 \mu_B$ measured for $\mathrm{UGe_2}$ \citenum{PfleidererPhysRevLett2002}), to the PM phase, as the hybridization magnitude $|V|$ increases. Here changing $|V|$ mimics its pressure variation. Both FM2$\rightarrow$FM1 and FM1$\rightarrow$PM transitions are of the first order as is observed for $\mathrm{UGe_2}$ below the critical end-point, though the FM2$\rightarrow$FM1 transition is of weak-first-order due to proximity to the quantum tricritical point \cite{WysokinskiPhysRevB2015} [see panel (d)]. Notably, our model also provides the value of magnetic moment $m \sim 1.6\,\mu_B$ in FM2 phase, close to the experimental $m \approx 1.45\,\mu_B$.\cite{PfleidererPhysRevLett2002}

\begin{figure}[h!]
   \centering
   \includegraphics[width = 1\columnwidth]{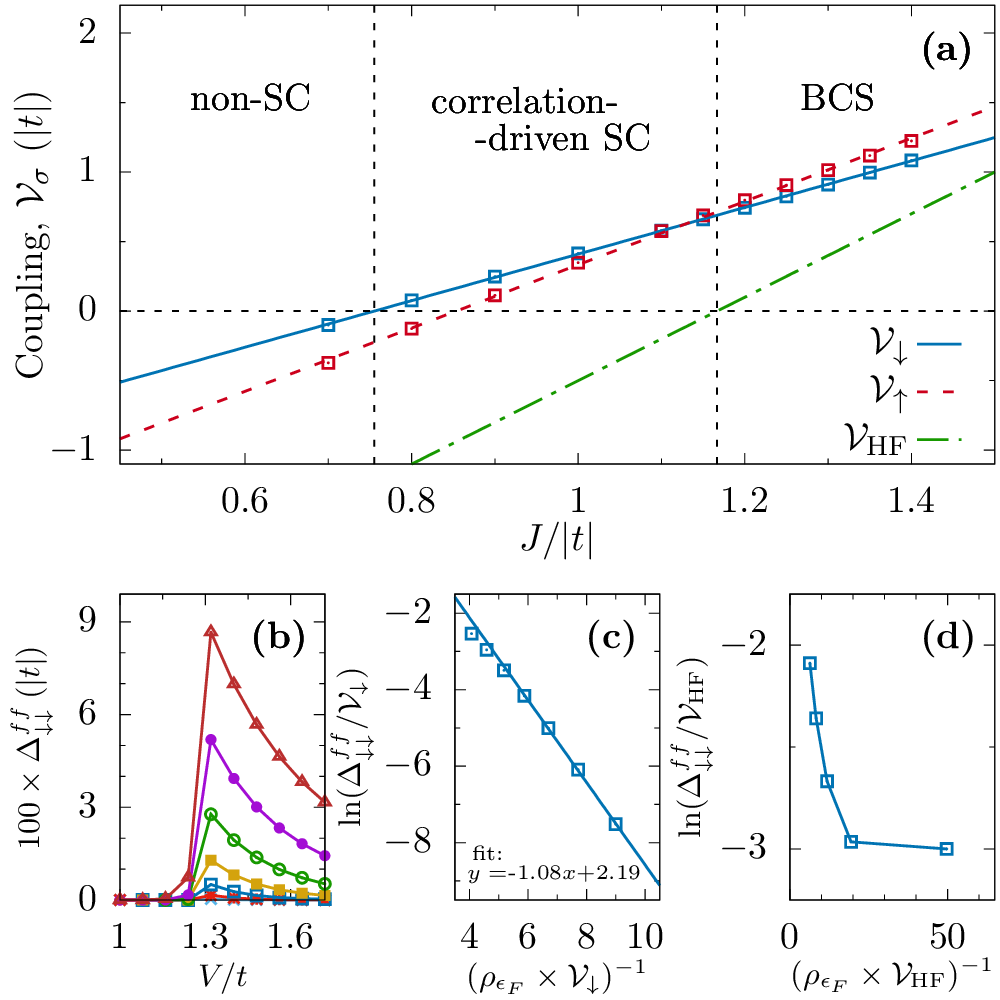}
     \caption{(a) Dependence of $\mathcal{V}_{\sigma}$ on the Hund's coupling $J$ for $V/t = 1.32$ (solid blue and dashed red lines). For comparison, the value of the Hartree-Fock (HF/BCS) coupling constant $\mathcal{V}_\mathrm{HF}$ is also shown by green dash-dotted line. Black dashed vertical lines split the plot into three regions: non-SC, correlation-driven (where SC is not supported at the HF/BCS level, yet is appears due to correlation effects), and BCS regime (where SC phase emerges in the HF/BCS approximation). Note that the value $J/|t| = 1.1$, considered above, falls into the correlation-driven regime. (b) Hybridization-dependence of the SC gap component $\Delta^{ff}_{\downarrow\downarrow}$ for various $J$ near the FM2$\rightarrow$FM1 transition. Values of $J/|t|$ (from top to bottom) are $1.4$, $1.35$, $1.3$, $1.25$, $1.2$, $1.15$, and $1.1$. (c) Scaling of $\Delta^{ff}_{\downarrow\downarrow}$ with the dimensionless effective coupling $\rho_{\epsilon_F} \mathcal{V}_\downarrow$. Here $\rho_{\epsilon_F}$ denotes total density of states per $f$-orbital per spin, evaluated at the Fermi energy in the normal phase. The gap follows renormalized BCS scaling $\Delta^{ff}_{\downarrow\downarrow} \propto \mathcal{V}_{\mathrm{\downarrow}} \times \exp(- (\rho_{\epsilon_F} \mathcal{V}_{\mathrm{\downarrow}})^{-1} )$. (d) The same as in (c), but with HF/BCS coupling $\mathcal{V}_\mathrm{HF}$ used instead of $\mathcal{V}_{\mathrm{\downarrow}}$. Breakdown of BCS scaling implies relevance of the correlation-driven coupling renormalization.}%
  \label{fig:bcs_scaling}
\end{figure}

The novel feature, inherent to the degenerate ALM and the principal result of present paper, is the emergence of distinct even-parity spin-triplet SC phases appearing around the magnetic transition points and characterized by non-zero SC gap parameters $\Delta_{\sigma\sigma}^{ff} \equiv \mathcal{V}_\sigma \langle \hat{f}_{i\sigma}^{(1)} \hat{f}_{i\sigma}^{(2)}\rangle_0$, as depicted in Fig.~\ref{fig:phase_diagram}(b). The $A_1$-type SC (i.e., the majority-spin gap $\Delta^{ff}_{\uparrow\uparrow} = 0$ and $\Delta^{ff}_{\downarrow\downarrow} \neq 0 $) sets in inside the FM1 phase and transforms to either $A_2$ phase ($\Delta^{ff}_{\downarrow\downarrow} > \Delta^{ff}_{\uparrow\uparrow} \neq 0$) at FM2-FM1 border or to $A$ state ($\Delta^{ff}_{\uparrow\uparrow} = \Delta^{ff}_{\downarrow\downarrow} \neq 0$) close to the FM1$\rightarrow$PM transition point. The latter two states appear in very narrow regions, as illustrated in Fig.~\ref{fig:phase_diagram}(e) and (f). The $A_2$-phase gap is by an order of magnitude smaller than its $A_1$ counterpart, whereas the $A$-phase gap is by even four orders of magnitude smaller. Hence, one can safely say that the $A_1$ phase is so far the only one observable for $\mathrm{UGe_2}$; the $A_2$ state could be detectable in applied magnetic field.\cite{Fidrysiak2unpublished} Note also that the pairing potential $\mathcal{V}_{\downarrow}$ is maximal near the corresponding metamagnetic transition [cf. Fig.~\ref{fig:phase_diagram}(c)]. Remarkably, this situation appears without any additional spin-fluctuation effect involved, which distinguishes the present mechanism from those invoked previously for the U-compounds.\cite{SandemanPhysRevLett2003,HattoriPhysRevLett108,TadaJPhysConfSeries2013} In the inset of Fig.~\ref{fig:phase_diagram}(b), we plot the specific-heat discontinuity (the shaded area) and the related magnetization jumps observed experimentally. The peaks identify the regime of bulk SC; these sharp features are reproduced by our calculation [cf. Fig.~\ref{fig:phase_diagram}(b)] and should be contrasted with the first resistivity data.\cite{SaxenaNature2000} Note also that we obtain small, but clear SC gap discontinuities at both $A_2 \rightarrow A_1$ and $A_1\rightarrow A$ transitions (cf. Fig.~\ref{fig:phase_diagram}(e) and (f), respectively). We emphasize that all the singularities are physically meaningful and well within the numerical accuracy (error bars are shown explicitly for the $A$ phase having the smallest gap magnitude). 

The nature of FM2 and FM1 phases can be understood by inspection of the corresponding spin- and orbital-resolved densities of states exhibited in Fig.~\ref{fig:dos}. In FM2 state [Fig.~\ref{fig:dos}(a)] $f$-electrons are close to localization and well below the Fermi energy $\epsilon_F$ as they carry out nearly saturated magnetic moments, whereas in FM1 phase [Fig.~\ref{fig:dos}(b)] $\epsilon_F$ is placed in the region of spin-down electrons, stabilizing the magnetization plateau (and illustrating the \textit{half-metallic character}), hence only $\Delta_{\downarrow \downarrow}^{ff} \neq 0$. Similar evolution of magnetism has been observed previously for the orbitally non-degenerate model.\cite{WysokinskiPhysRevB2014,WysokinskiPhysRevB2015,AbramJMagnMagnMat2016} Fig.~\ref{fig:dos}(c) illustrates the paramagnetic behavior.

\begin{figure}
  \includegraphics[width = \columnwidth]{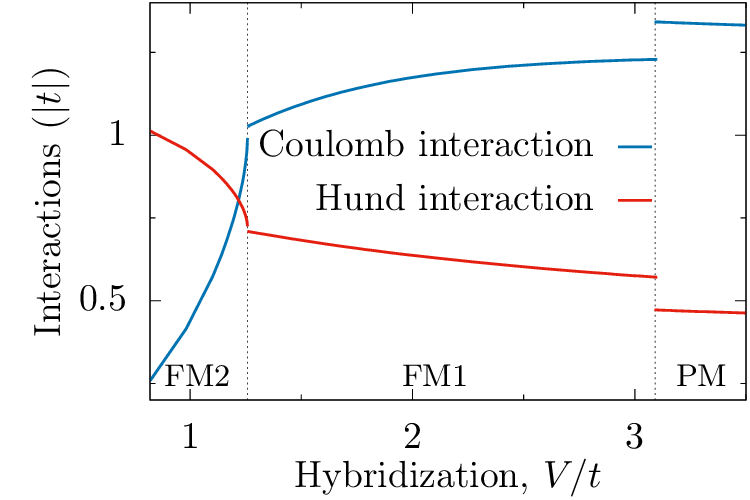} %
  \caption{Comparison of the intra-orbital Coulomb  and Hund's rule coupling   contributions to the total system energy as a function of the hybridization magnitude $V/t$. These two quantities are defined as $2U \left< \hat{n}^{f(1)}_{i \uparrow}  \hat{n}^{f(1)}_{i \downarrow} \right>_{\mathrm{G}}$ and $2J \left<  \left( \mathbf{\hat{S}}_i^{f(1)} \cdot \mathbf{\hat{S}}_i^{f(2)} + \frac{1}{4} \hat{n}_i^{f(1)} \hat{n}_i^{f(2)} \right)\right>_{\mathrm{G}}$, respectively. The two energies are comparable near the FM2$\rightarrow$FM1 borderline, hence called the Hund's-metal regime (see the main text).}%
  \label{fig:hunds_metal}
\end{figure}

Next, we discuss the fundamental role of the effective pairing potential. Explicitly, in Fig.~\ref{fig:bcs_scaling}(a) we have plotted renormalized and bare coupling constants as a function of $J$ for $V/t = 1.32$. The dominant component $\mathcal{V}_\downarrow$ remains positive down to $J/|t| \approx 0.76$, whereas the HF/BCS coupling changes sign already for $J/|t| = 3.5/3 \approx 1.17$. Electronic correlations are thus the crucial factor stabilizing the triplet SC close to the FM2-FM1 boundary. Fig.~\ref{fig:bcs_scaling}(b) shows the dominant gap component for selected values of $J$. The gap increases very rapidly with the increasing Hund's rule coupling, as detailed in Fig.~\ref{fig:bcs_scaling}(c), where we plot logarithm of the normalized gap, $\ln (\Delta_{\downarrow\downarrow}^{ff} / \mathcal{V}_\downarrow)$ vs $(\rho_{\epsilon_F} \mathcal{V}_\downarrow)^{-1}$ for fixed hybridization $V/t = 1.32$, that corresponds to the $A_1$ phase ($\rho_{\epsilon_F}$ is the total density of states per $f$-orbital per spin at $\epsilon_F$). A good linear scaling is observed with the coefficient $\approx -1.08$, not to far from the BCS value $-1$. The binding of $f$-electrons into local triplet pairs is provided partly by the Hund's rule exchange that yields the HF/BCS potential $\mathcal{V}_\mathrm{HF} = 3 J - U$. Fig.~\ref{fig:bcs_scaling}(d) shows the same as Fig.~\ref{fig:bcs_scaling}(c), but $\mathcal{V}_\mathrm{HF}$ has been taken in place of $\mathcal{V}_\sigma$. The breakdown of the scaling implies there a significant effect of local correlations over the Hund's-rule induced pairing. The relevance of the local Coulomb interactions combined with the Hund's rule physics can be also seen by comparing the contributions the intra-orbital Coulomb-repulsion and inter-orbital Hund's rule coupling to the total ground-state energy (Fig.~\ref{fig:hunds_metal}). Close to the metamagnetic FM2$\rightarrow$FM1 transition, where the SC amplitude is the largest, those two scales are comparable and of the order of the kinetic term $|t|$. This places the system in the correlated \textit{Hund's metal regime}, previously coined in the context of Fe-based SC.\cite{YinNature2011}

\section{Discussion and conclusions}
\label{sec:discussion}

To underline the quantitative aspect of our analysis of the SC phase we have determined the temperature dependence of the gap in the combined FM1$+A_1$ state for $J/|t| =  1.1$ and $V/t = 1.3$, i.e., near the gap-maximum point depicted in Fig.~\ref{fig:phase_diagram}(b). Selecting the value of $|t| = 0.5\,\mathrm{eV}$, we obtain SC critical temperature $T_\mathrm{SC} \approx 0.92\,\mathrm{K}$ (see Appendix~\ref{appendix:non_zero_temperature}), very close to the experimental value $T_\mathrm{SC} \sim 0.75\,\mathrm{K}$ in the highest-quality samples.\cite{HaradaPhysRevB2007} Note that for $J \lesssim 1.17 |t|$ we do not expect any SC in the HF/BCS approximation. It is gratifying that the value of $J = 1.1|t| = 0.55\,\mathrm{eV}$ can lead to such a subtle SC temperature scale $T_\mathrm{SC} < 1\,\mathrm{K}$ in the situation, where the FM transition temperature $T_c$ is by two orders of magnitude larger or even higher. Equally important is the obtained value of specific-heat jump $\Delta C/(\gamma_n T_\mathrm{SC}) \simeq 1.44$ (cf. Fig.~\ref{fig:specific_heat}), i.e., very close to the BCS value 1.43. Parenthetically, this is not too far from experimental $\Delta C/(\gamma_n - \gamma_0) / T_\mathrm{SC} \simeq 0.97$ for pressure $1.22\,\mathrm{GPa}$ \cite{TateiwaPhysRevB2004} (corresponding closely to our choice of parameters) if we subtract the residual Sommerfeld coefficient $\gamma_0$.

\begin{figure}
	\includegraphics[width = \columnwidth]{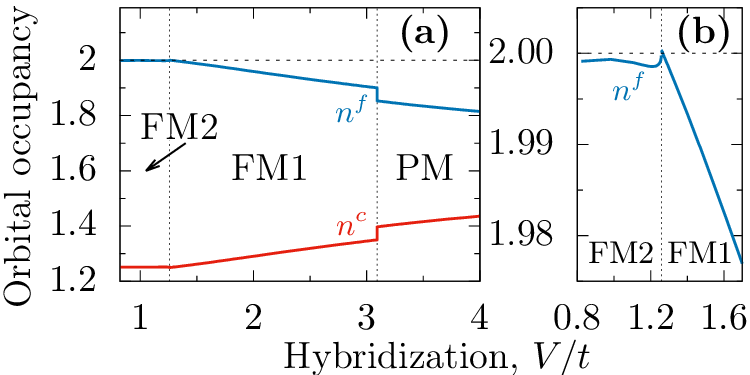} %
	\caption{(a) Occupancies of the $f$- and $c$-orbitals as a function of hybridization. (b) Close-up of the FM2$\rightarrow$FM1 transition.}%
	\label{fig:occupancies}
\end{figure}

The $\mathrm{U^{3+}}$ ionic configuration is $5f^3$. Some experimental evidence points to the value close to $\mathrm{U^{4+}}$ ($5f^2$).\cite{PfleidererRevModPhys2009,TrocPhysRevB2012} Here the good values of magnetic moments in both FM2 and FM1 phases are obtained for approximate $5f^2$ configuration and $n^c \approx 1.25$ conduction electrons, as shown in Fig.~\ref{fig:occupancies}(a). Namely, the results in Fig. \ref{fig:occupancies}(b) point clearly to the value $n^f \approx 2$ in FM2 phase and it diminishes almost linearly in FM1 state. Such a behavior explains that the two \emph{f}-electrons are practically localized in the FM2 phase and therefore, no SC state induced by the Hund's rule and \emph{f}-\emph{f} correlations can be expected. On the other hand, the correlations are weaker on the PM side due to substantially larger hybridization and, once again, SC disappears. These results suggest that here the third \emph{f}-electron may have become selectively itinerant and thus is weakly correlated with the remaining two. It is tempting to ask about its connection to the residual value of $\gamma_0$ at $T\rightarrow0$ and to $m_0$ for $T > T_c$.

In summary, our theoretical phase diagram reproduces the fundamental features observed experimentally in a semiquantitative manner. Within the double-degenerate Anderson lattice model in the statistically consistent renormalized mean-field approximation (SGA), we have analyzed in detail the coexisting FM1 and spin-triplet $A_1$ SC phase, having in mind the experimental results for $\mathrm{UGe_2}$. We obtain also an indirect evidence for an orbital-selective Mott-type delocalization of one of $5f$ electrons at low temperature, which may be followed by its gradual localization in the high-temperature (Curie-Weiss) regime, leading to $\mathrm{U}^{3+}$ magnetic configuration as exhibited by static magnetic susceptibility.  Further specific material properties of $\mathrm{UGe_2}$ and related systems can be drawn by incorporating the angular dependence of the hybridization, more realistic multi-orbital structure, as well as the third-dimension.

It would be interesting to incorporate, \textit{renormalized} in this situation, quantum spin fluctuations into our SGA (renormalized-mean-field-theory-type picture). Such an approach would start from the effective Landau functional for fermions $\mathcal{F}$. (cf. Appendix~\ref{appendix:SGA}) and a subsequent derivation of the corresponding functional involving magnetic-moment fluctuations as an intermediate step, that would allow to include their contribution to the resultant free energy. Such a step, if executed successfully, would represent a decisive step beyond either the spin-fluctuation or the real-space-correlation approach. We should be able to see progress along these lines in near future.

This work was supported by MAESTRO Grant No. DEC-2012/04/A/ST3/00342  from  Narodowe  Centrum  Nauki (NCN). 

\appendix

\section{Statistically-consistent Gutzwiller approximation (SGA)}
\label{appendix:SGA}

Here we present technical details of the Statistically Consistent Gutzwiller Approximation (SGA), as applied to the four-orbital model discussed in the main text. At zero temperature, this variational technique reduces to the problem of  minimizing  the energy functional $E_G = \langle\Psi_G| \mathcal{H} |\Psi_G\rangle / \langle\Psi_G|\Psi_G\rangle$ with respect to the trial state $|\Psi_G\rangle = P_G |\Psi_0\rangle$ for fixed electron density. $|\Psi_0\rangle$ is (\textit{a priori} unknown) wave-function describing  Fermi sea of free quasi-particles, whereas $\hat{P}_G = \prod_{il} \hat{P}^{(l)}_{G i}$ denotes Gutzwiller correlator.\cite{GutzwillerPhysRevLett1963} Local correlators  $\hat{P}^{(l)}_{Gi} = \sum_\alpha \lambda_{\alpha} |l \alpha\rangle_i {}_i\langle l\alpha|$ adjust weights of configurations  $\alpha \in \{\emptyset, \uparrow, \downarrow, \uparrow\downarrow\}$ on each $f$-orbital (indexed by $l$) at site $i$ by means of coefficients $\lambda_\alpha$ multiplying projection operators $|l \alpha\rangle_i {}_i\langle l\alpha|$. This is not the most general form of $\hat{P}_G$,\cite{KubiczekMasterThesis} but generalization makes the results less transparent and leads only to minor numerical corrections which may be safely disregarded. Evaluation of the expectation values with the correlated wave function is a non-trivial many-body problem. The latter can be substantially simplified by setting up a formal expansion about the limit of infinite lattice coordination,  which is achieved by imposing a constraint $(\hat{P}^{(l)}_{Gi})^2 \equiv 1 + x \times \Pi_\sigma (\hat{n}^{f(l)}_{i\sigma} - n^{f(l)}_\sigma)$ \cite{BuenemannEurophysLett2012} so that all $\lambda_\alpha$ are now expressed in terms of single variational parameter $x$ (we have introduced the notation $O \equiv \langle \hat{O}\rangle_0 \equiv \langle\Psi_0|\hat{O}|\Psi_0\rangle$ for general operator $\hat{O}$). This approach has been elaborated in detail earlier for the orbitally-degenerate Hubbard and non-degenerate Anderson model \cite{WysokinskiPhysRevB2014,AbramJMagnMagnMat2016,AbramJMagnMagnMat2016,ZegrodnikNewJPhys2013, ZegrodnikNewJPhys2014}.

We now focus on the four-orbital model, discussed in the text, and calculate  $E_G$ by means of Wick theorem, allowing for non-zero equal-spin pairing amplitudes $\langle \hat{f}^{(1)}_{i\sigma} \hat{f}^{(2)}_{i\sigma} \rangle_0$, orienting magnetization direction along $z$ axis, and resorting to the Gutzwiller approximation by discarding the contributions irrelevant for infinite lattice coordination. In effect, we obtain

\begin{align}
E_G \simeq   &\sum_{i j l \sigma} t_{ij} \langle\hat{c}^{(l)\dagger}_{i\sigma} \hat{c}^{(l)}_{j\sigma}\rangle_0 + V \sum_{i l \sigma} q_{\sigma} \left( \langle\hat{f}^{(l)\dagger}_{i\sigma} \hat{c}^{(l)}_{i\sigma}\rangle_0 + \mathrm{C.c.}\right) \nonumber \\ + & \sum_{i \sigma}\left[U'  g_{1\sigma} + (U' - J)  g_{2\sigma}\right] |\langle\hat{f}^{(1)}_{i\sigma} \hat{f}^{(2)}_{i\sigma}\rangle_0|^2  + \nonumber \\ & \sum_i \left[-2J S^{z f(1)}_i S^{z f(2)}_i + (U' - \frac{J}{2}) n^{f(1)}_i n^{f(2)}_i\right]  + \nonumber\\ & + \epsilon^f \sum_{i l} n^{f(l)}_i + U  \sum_{il} \lambda_{\uparrow \downarrow}^2 n^{f(l)}_{i\uparrow} n^{f(l)}_{i\downarrow}, 
\end{align}
where the renormalization factors are defined as
\begin{align}
  \label{eq:renormalization_factors}
 q_\sigma =& \lambda_\emptyset \lambda_\sigma + (\lambda_{\uparrow\downarrow} \lambda_{\bar{\sigma}} - \lambda_\emptyset \lambda_\sigma) \times n^{f(l)}_{\bar{\sigma}}, \nonumber \\ g_{1\sigma} = & 2 \times (\lambda_{\uparrow\downarrow}^2 - \lambda_{\bar{\sigma}}^2)\times(\lambda_\sigma^2 + (\lambda_{\uparrow\downarrow}^2 - \lambda_\sigma^2)  n^{f(l)}_{\bar{\sigma}})\times n^{f(l)}_{\bar{\sigma}}, \nonumber \\  g_{2\sigma} = & (\lambda_{\uparrow\downarrow}^2 - \lambda_{\bar{\sigma}}^2)^2 \times \left(n^{f(l)}_{\bar{\sigma}}\right)^2 + (\lambda_\sigma^2 + (\lambda_{\uparrow\downarrow}^2 -
\lambda_\sigma^2)  n^{f(l)}_{\bar{\sigma}})^2. 
\end{align}

The SGA method maps the original many-body problem onto the task of calculating an effective Landau functional $\mathcal{F} = -\beta^{-1} \ln \mathrm{Tr} \exp(-\beta \mathcal{H}_\mathrm{eff})$ evaluated with the effective one-body Hamiltonian $\mathcal{H}_\mathrm{eff} = E_G(\{P_\gamma, x\}) - \mu N_e + \sum_\gamma \lambda_\gamma (\hat{P}_\gamma - P_\gamma )$, where $N_e$ is total number of electrons in the system, $\gamma$ runs over bilinears $\hat{P}_\gamma$ composed of creation and annihilation operators, and $\lambda_\gamma$ are Lagrange multipliers ensuring that $P_\gamma$ obtained from optimization of $\mathcal{F}$ and the Bogolubov-de Gennes equations coincide. The values of parameters are determined from the equations  $\partial_{P_\gamma} \mathcal{F} = 0$, $\partial_{x} \mathcal{F} = 0$, and $\partial_{\lambda_\gamma} \mathcal{F} = 0$. Additionally, the value of chemical potential $\mu$ is fixed by electron density. Note that the original variational problem is well posed at $T = 0$, whereas the SGA formulation is applicable also for $T > 0$. One can argue (for general coordination number) that for $T \rightarrow 0$ optimization of $\mathcal{F}$ with $\mathcal{H}_\mathrm{eff}$ yields the variational minimum of $E_G$ within the improved Gutzwiller approximation,\cite{KaczmarczykNewJPhys2014} whereas for $T > 0$ it reflects thermodynamics of projected quasi-particles. \cite{Fidrysiak_unpublished}

Explicit form of the effective Hamiltonian reads
\begin{align}
	\mathcal{H}_\mathrm{eff}  = \sum_{\mathbf{k}, \sigma}^{} \Psi_{\mathbf{k} \sigma}^\dagger \left( \begin{array}{cccc}
		\epsilon_{ \mathbf{k} } & 0 &  q_\sigma V&  0\\
		0 & - \epsilon_{ \mathbf{k} } & 0 & - q_\sigma V\\
		q_\sigma V & 0 & \epsilon_{ \sigma }^{f} & \Delta_{ \sigma \sigma }^{ff} \\
		0 & - q_\sigma V & \Delta_{ \sigma \sigma }^{ff} & -\epsilon_{ \sigma }^{f} \\
	\end{array}
	\right) \Psi_{\mathbf{k} \sigma}
  + E_0,
  \label{eq:effective_hamiltonian}
\end{align}
where $\Psi_{\mathbf{k}\sigma }^\dagger = \left( \hat{c}^{(1)\dagger}_{ \mathbf{k} \sigma }, \hat{c}^{(2)}_{ -\mathbf{k} \sigma }, \hat{f}^{(1)\dagger}_{ \mathbf{k} \sigma }, \hat{f}^{(2)}_{ -\mathbf{k} \sigma } \right)$, $\epsilon_{ \mathbf{k} } =  2 t [\cos(k_x) + \cos(k_y)] + 4t^\prime \cos(k_x)\cos(k_y) - \mu$ is the conduction band dispersion,
\begin{align}
\Delta_{ \sigma \sigma }^{ff} = [g_{1\sigma} U^\prime + g_{2\sigma} (U^\prime - J)] \times  \langle\hat{f}^{(1)}_{i\sigma} \hat{f}^{(2)}_{i\sigma}\rangle_0   
\end{align}
denotes $f$-$f$ superconducting gap parameter,
        	\begin{align}
		\epsilon_{ \sigma }^f &= \frac{\partial E_G}{\partial n^{f(1)}_{i \sigma}}  =                  
		\epsilon^f + U \lambda_{\uparrow \downarrow}^2  n^{f(1)}_{i \bar{\sigma}}+ ( U^\prime - J ) n^{f(2)}_{i \sigma} + U^\prime n^{f(2)}_{i \bar{\sigma}} + \nonumber \\&
		+ 
		\left( \frac{\partial q_{\bar{\sigma}}}{\partial n^{f(1)}_{i \sigma}} V \sum_l\langle\hat{f}^{(l)\dagger}_{i\bar{\sigma}} \hat{c}^{(l)}_{i\bar{\sigma}}\rangle_0 + \mathrm{C.c.} \right)	
		+ \nonumber \\ &+
		\left( \frac{\partial g_{1\bar{\sigma}}}{\partial n^{f(1)}_{i \sigma}} U^\prime + \frac{\partial g_{2\bar{\sigma}}}{\partial n^{f(1)}_{i \sigma}} (U^\prime - J) \right) |\langle\hat{f}^{(1)}_{i\bar{\sigma}} \hat{f}^{(2)}_{i\bar{\sigma}}\rangle_0|^2 - \mu 
	\end{align}
        is the renormalized \emph{f}-orbital energy, and $E_0 \equiv E_G(\{P_\gamma, x\}) - \mu N_e - \sum_\gamma \lambda_\gamma P_\gamma$ is a  remainder proportional to unity. Note that the entries of $\mathcal{H}_\mathrm{eff}$ have been obtained from one condition $\partial_{P_\gamma}\mathcal{F} = 0$ and are given in an explicit form.

\begin{widetext}
        
Since the effective Hamiltonian \eqref{eq:effective_hamiltonian}  can be diagonalized analytically, with the eigenvalues

	\begin{align}
	E_{ \mathbf{k} \sigma }^{(\lambda)} = \pm \sqrt{
	q_\sigma^2 V^2 
	+ \frac{1}{2} \left[ \left( \Delta^{ff}_{ \sigma \sigma } \right)^2 + \left( \epsilon^f_\sigma \right) ^2 + \epsilon_\mathbf{ k }^2 \right]
	\pm \frac{1}{2} \sqrt{
		\left[
			\left( \Delta^{ff}_{ \sigma \sigma } \right)^2 + \left( \epsilon^f_\sigma \right)^2 - \epsilon_\mathbf{ k }^2 
		\right]^2 
		+ 4 q_\sigma^2 V^2   	
		\left[
		\left( \Delta^{ff}_{ \sigma \sigma } \right)^2 
		+ 
		\left(
		\epsilon_\mathbf{ k }
		+ \epsilon^f_\sigma
		\right)^2  
		\right]} 
	}, 
	\end{align}

      \noindent
one can express the gap $\Delta_\mathbf{k}$ in the projected quasi-particle spectrum in terms of the gap parameter $\Delta_{\sigma\sigma}^{ff}$. We get the formula

\begin{align}
\Delta_\mathbf{k}^2 = \frac{\epsilon_\mathbf{k}^2}{(\epsilon_\mathbf{k} + \epsilon_\sigma^f)^2} \times (\Delta_{\sigma\sigma}^{ff})^2 + o[(\Delta_{\sigma\sigma}^{ff})^2],
\end{align}

\noindent
valid for wave vectors located on the Fermi surface calculated in the normal state. Note that the gap is expressed solely in terms of the $f$-$f$ pairing amplitude (even though $f$-$c$ and $c$-$c$ amplitudes are, in general, non-zero due to the hybridization effects) and scaled by the $\mathbf{k}$-dependent factor. This justifies using $\Delta_{\sigma\sigma}^{ff}$ as the quantity characterizing the overall SC properties of the system.

      \end{widetext}
        
\section{Numerical procedure}
\label{appendix:numerical_procedure}

The system of equations $\partial_{P_\gamma} \mathcal{F} = 0$, $\partial_{x} \mathcal{F} = 0$, and $\partial_{\lambda_\gamma} \mathcal{F} = 0$ has been solved by means of GNU Scientific Library. Numerical accuracy for the dimensionless density matrix elements has been chosen in the range $10^{-8}$-$10^{-9}$, depending on the model parameters. We work in the thermodynamic limit with number of lattice sites $N \rightarrow \infty$ by performing Brillouin-zone integration in all equations. Technically, keeping $N$ finite, but large speeds up the calculations in a highly parallel setup. However, the calculated superconducting gap parameters  range from $\sim 10^{-4}|t|$ down to $\sim 10^{-9}|t|$ which raises the question of the impact of the finite-size effects on the SC state. We can estimate the latter by referring to the Anderson criterion \cite{AndersonJPhysChemSolids1959} $\Delta_{\sigma\sigma}^{ff} \sim d$, where $d \sim W/N$ is the typical spacing between discrete energy levels ($W \sim$ several $|t|$ denotes bandwidth scale and $N$ is the number of lattice sites). To achieve the desired accuracy, one would thus need to consider lattices with $> 10^{10}$ sites. This rationalizes our choice to use adaptive integration and work directly with infinite system.

\begin{figure}
  \includegraphics[width = \columnwidth]{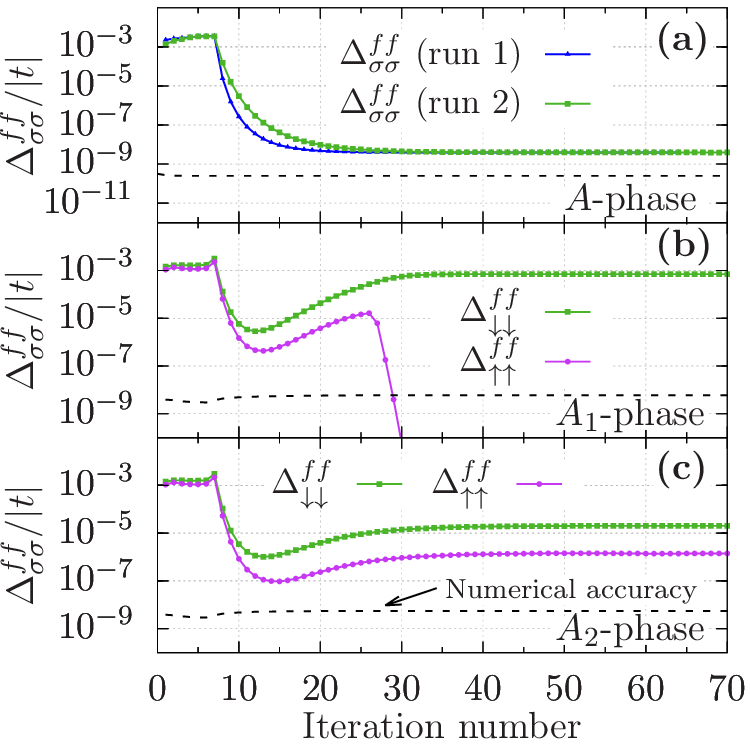} %
  \caption{Convergence properties of the solutions for the $A$, $A_1$, and $A_2$ SC phases as a function of the iteration number. The values of hybridization are: (a) $V/t \approx 3.0982$ ($\mathrm{PM}+A$-phase), (b) $V/t \approx 1.264$ ($\mathrm{FM1}+A_1$-phase just above $\mathrm{FM2}+A_2\rightarrow\mathrm{FM1}+A_1$ transition), and (c) $V/t \approx 1.254$ ($\mathrm{FM2}+A_2$-phase just below $\mathrm{FM2}+A_2\rightarrow\mathrm{FM1}+A_1$ transition). The remaining parameters are the same as in the main text. Panel (a) shows two independent runs for the equivalent $\Delta^{ff}_{\uparrow\uparrow} = \Delta^{ff}_{\downarrow\downarrow}$ gap components. In panels (b)-(c) the single run is displayed for two (inequivalent) gap parameters. The horizontal dashed lines mark the numerical accuracy.}%
  \label{fig:convergence}
\end{figure}

The convergence properties of our computational scheme for the parameters corresponding to the $A$-, $A_1$-, and $A_2$-phases are summarized in Fig.~\ref{fig:convergence}(a)-(c). Only the SC amplitudes are displayed (connected points); the dashed lines mark the target numerical accuracy set in our code (note that the accuracy varies at the initial stage of the procedure due to drift of the renormalization factors). In each case, we have performed a few warm-up iterations with imposed non-zero values of the SC gap parameters, symmetry-breaking external magnetic field, and finite temperature. This initial phase is seen in Fig.~\ref{fig:convergence} as a plateau for less then 10 iterations. Subsequently, the auxiliary fields were turned off and the system was allowed to relax. For the $A$-phase (zero total magnetization and $\Delta^{ff}_{\uparrow\uparrow} = \Delta^{ff}_{\downarrow\downarrow}$), and for the smallest gap amplitudes, we have executed the iterative procedure in different set-ups multiple times to verify the solutions. Two runs are marked in Fig.~\ref{fig:convergence}(a) by blue and green colors. In panels (b)-(c) green and purple lines show (now inequivalent) spin-down and spin-up amplitudes for the $A_1$ and $A_2$ phases. The ``S''-shaped iteration-dependence of the $\Delta^{ff}_{\uparrow\uparrow}$ in panel (b) may be attributed to the proximity to the first order $\mathrm{FM2}+A_2\rightarrow\mathrm{FM1}+A_1$ transition. The initial upturn of $\Delta^{ff}_{\uparrow\uparrow}$ is reminiscent of the behavior observed for the $A_2$ state [cf. panel (c)]. After $25$th iteration, however, the system switches to another attractor and $\Delta^{ff}_{\uparrow\uparrow}$ is exponentially suppressed to quickly attain the numerical zero ($A_1$ phase).

\begin{figure}
  \includegraphics[width = \columnwidth]{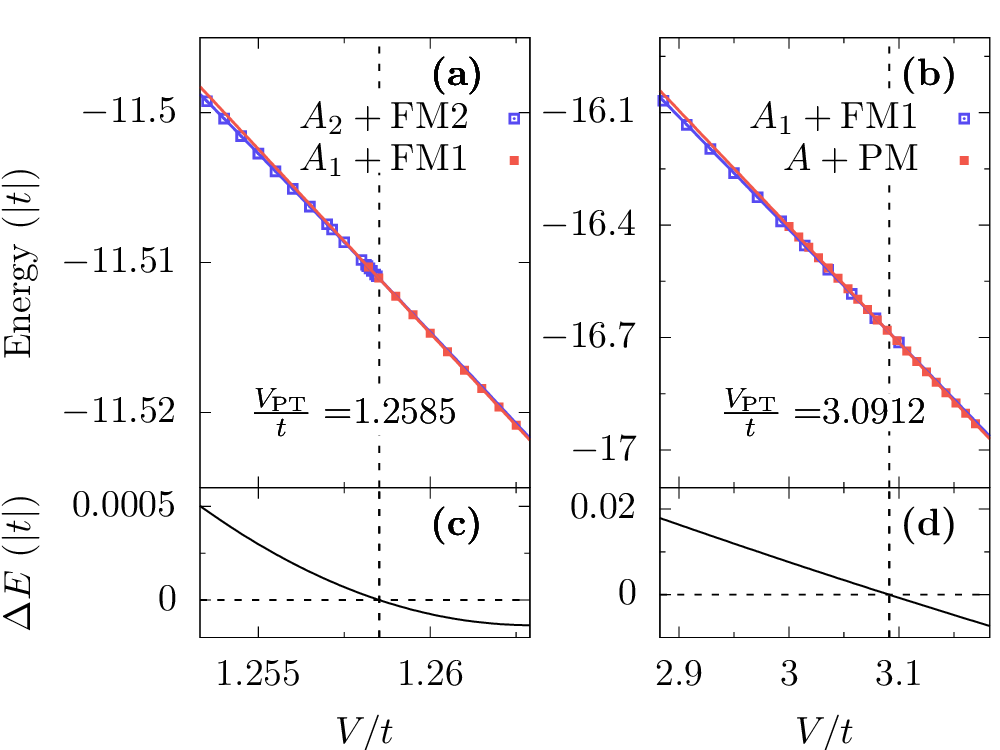} %
  \caption{Crossing of the energies near (a) $\mathrm{FM2}+A_2\rightarrow\mathrm{FM1}+A_1$ and (b) $\mathrm{FM1}+A_1\rightarrow \mathrm{PM}+A$ transition for $U / |t|= 3.5$. Panels (c)-(d) show the energy differences $\Delta E$ between extrapolated energies on both sides of respective phase transitions. The latter become zero at the transition point, denoted as $V_\mathrm{PT}$, and are displayed in the plot. Model parameters coincide with those used in Fig.~1. text.}%
  \label{fig:energy_crossing}
\end{figure}

\section{Determination of the phase diagram}
\label{appendix:phase_diagram}

In Fig.~\ref{fig:energy_crossing}(a) we plot the energies of the $\mathrm{FM2}+A_2$ and $\mathrm{FM1}+A_1$ phases near the metamagnetic transition for $J/|t| = 1.1$, $U/|t| = 3.5$, $t'/|t| = 0.25$, and $\epsilon^f / |t|= -4$ (the same parameters have been used for plotting Fig.~1). The solid lines are quadratic fits to the data in respective phases. The phase-transition point $V_\mathrm{PT}$ corresponds to the crossing of the lines (marked by the vertical dashed lines) and is displayed in the figure. Note that the lines cross at non-zero angle which is indicative of the first-order transition. Similarly, in Fig.~\ref{fig:energy_crossing}(b) the energies near the $\mathrm{FM1}+A_1$ and $\textrm{PM}+A$ phase boundary are shown. In panels (c)-(d) we plot the difference between extrapolated energies on both sides of the transitions. The latter becomes zero at the transition point.

For the sake of completeness, in Table~\ref{tab:energies} we present the analysis of $A_1$-type SC phase stability for $U/|t| = 3.5$, $V/t = 1.32$, $t'/|t| = 0.25$,  $\epsilon^f / |t| = -4$, and variable Hund's coupling $J/|t| = 1.1 \div 1.4$. Here $E_\mathrm{FM1}$ is the energy of the FM1 phase with SC suppressed, and $E_{\mathrm{FM1}+A_1}$ refers to the FM1 phase coexisting with $A_1$-type SC. The condensation energy $E_c \equiv E_\mathrm{FM1} - E_{\mathrm{FM1}+A_1}$ is positive for all considered values of hybridization, which illustrates the stable character of the SC state.

\begin{table}
  \centering
  \caption{Variational ground-state energies for $U/|t| = 3.5$ and $V/t = 1.32$, $t'/|t| = 0.25$, $\epsilon^f/|t| = -4$, $n^\mathrm{tot} = 3.25$, and selected values of Hund's coupling $J$.  Here $E_\mathrm{FM1}$ is the energy of the FM1 phase with SC suppressed, and $E_{\mathrm{FM1}+A_1}$ refers to the FM1 phase coexisting with the $A_1$-type SC. The condensation energy $E_c \equiv E_\mathrm{FM1} - E_{\mathrm{FM1}+A_1}$ is also supplied. The numerical accuracy of the energy difference is of the order of $2 \times 10^{-8}$.}
  \label{tab:energies}
  
  \begin{ruledtabular}
  \begin{tabular}{ c  c  c  c}
  $J/|t|$ & $E_\mathrm{FM1}/|t|$ & $E_{\mathrm{FM1}+A_1}/|t|$ & $10^4 \times E_c/|t|$ \\ 
	\hline
1.10 & -11.663 459 37   & -11.663 459 39 &  0.0003 \\ 
1.15 & -11.796 917 55   & -11.796 917 77 & 0.0022 \\  
1.20 & -11.934 039 49   & -11.934 041 79  & 0.0230 \\ 
1.25 & -12.074 935 82   & -12.074 951 80 & 0.1598 \\ 
1.30 & -12.219 724 82   & -12.219 802 79 &  0.7797 \\ 
1.35 & -12.368 534 05   & -12.368 822 15 & 2.8810 \\ 
1.40 & -12.521 501 70   & -12.522 354 69 &  8.5299 \\
\end{tabular}
\end{ruledtabular}
\end{table}

\section{Nonzero-temperature properties}\
\label{appendix:non_zero_temperature}

\begin{figure}
  \includegraphics[width = \columnwidth]{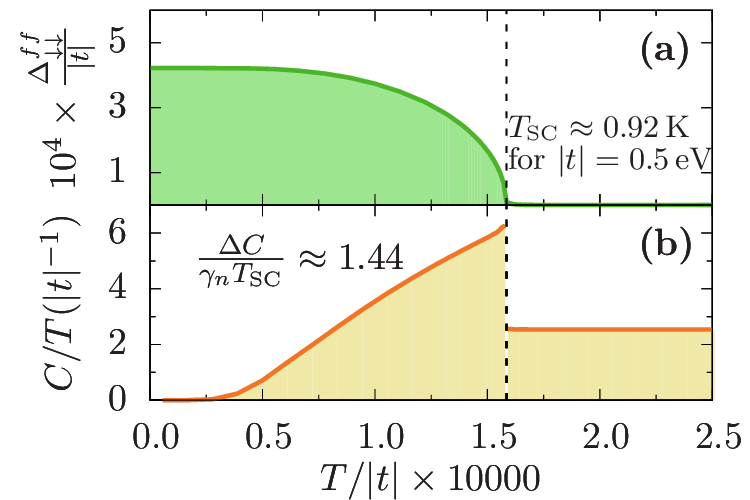} %
\caption{Temperature-dependence of (a) gap parameter $\Delta_{\downarrow\downarrow}^{ff}$ and (b) electronic specific heat for  $U / |t| = 3.5$, $J/|t| = 1.1$, $V / t = 1.3$, $\epsilon^f/|t| = -4$, $t'/|t| = 0.25$, and $n^\mathrm{tot} = 3.25$.}%
  \label{fig:specific_heat}
\end{figure}

Within the SGA approach, one can also determine the finite-temperature properties of the system. In Fig.~\ref{fig:specific_heat} we show explicitly the evolution of the gap parameter $\Delta_{\downarrow\downarrow}^{ff}$ and electronic specific heat across the SC transition for $U/|t| = 3.5$, $J/|t| = 1.1$, $V / t = 1.3$, $t'/|t| = 0.25$, and $\epsilon^f = -4$. For this set of parameters the system is close to the FM2$\rightarrow$FM1 transition, where SC is most pronounced (cf. Fig.~1). For the specific choice $|t| = 0.5\,\mathrm{eV}$ we obtain the SC transition temperature $T_\mathrm{SC} \simeq 0.92\,\mathrm{K}$ which is close to the values measured for high-quality $\mathrm{UGe_2}$ samples. On the other hand, we do not get the residual $C/T$ for $T\rightarrow 0$ as is observed for $\mathrm{UGe_2}$. This is likely due to more complex electronic structure, not included in the minimal four-orbital model considered here, e.g., by the third $5f$-electron, which provides the orbital-selective delocalized state, as discussed in the text. This conjecture is substantiated by the fact that if we subtract the residual $\gamma_0$ from measured Sommerfeld coefficient $\gamma_n$ then $\Delta C/(\gamma_n - \gamma_0)/T_\mathrm{SC} \simeq 0.97$,\cite{TateiwaPhysRevB2004} i.e., not too far from the value displayed in Fig.~\ref{fig:specific_heat}(b), which, in turn, is close to the BCS value 1.43. \cite{KishoreEurPhysJB1999}

\section{Phase diagram in the regime of large Hund's coupling}
\label{appendix:large_hunds_coupling}

\begin{figure}
  \includegraphics[width = \columnwidth]{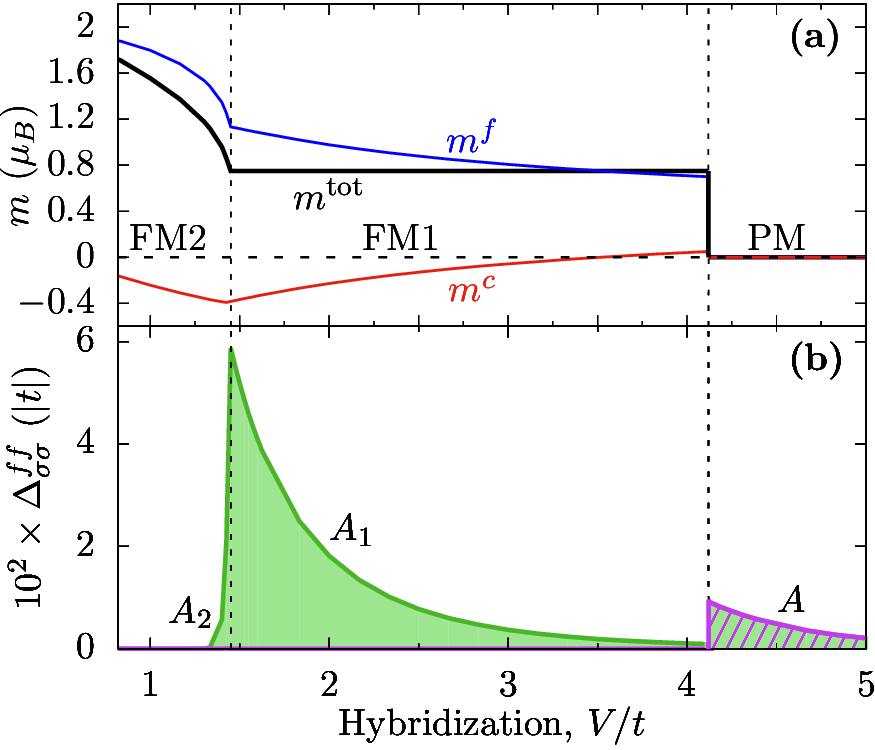} %
\caption{Phase diagram for $U / |t| = 4$, $J / |t| = 1.6$, $t'/|t| = 0.25$, temperature $T / |t| = 10^{-8}$ and $n^\mathrm{tot} = 3.25$. (a) Total magnetic moment (black line) and $f$- and $c$-electron magnetizations (blue and red lines, respectively). (b) Superconducting gap parameters $\Delta_{\downarrow\downarrow}^{ff}$ (green shading) and $\Delta_{\uparrow\uparrow}^{ff}$ (purple shading).}%
  \label{fig:phase_diagram_J160}
\end{figure}

For the parameters taken in the main text, the $A$-phase gaps turn out to be of the order $\Delta^{ff}_{\sigma\sigma} / |t| \sim 10^{-9}$, which  sets the critical temperature scale at the level of $0.01\,\mathrm{mK}$ for $|t| \sim 1\,\mathrm{eV}$. This raises a question about, limited to special situations, observability of the $A$ state. Here we show that the $A$ phase may become substantially enhanced in the regime of strong correlations and large  Hund's coupling. In Fig.~\ref{fig:phase_diagram_J160} we show the hybridization-dependence of the magnetization and SC gaps for $U / |t| = 4$, $J / |t| = 1.6$, $t'/|t| = 0.25$, $\epsilon^f / |t| = -4$, and temperature $T / |t| = 10^{-8}$. The general structure of the phase diagram remains unchanged, but the ratio of the gap parameters in the $A$ and $A_1$ phases is now enhanced by five orders of magnitude relative to the situation considered previously as that corresponding to the $\mathrm{UGe_2}$ case. However, this last feature suggests that an $A$-like phase could emerge in systems more strongly correlated than $\mathrm{UGe_2}$. Also, now the $A_1$ phase is not concentrated in a narrow region around the metamagnetic transition, but spreads over the entire FM1 region of the phase diagram. This is not consistent with the low-temperature specific-heat data \cite{TateiwaPhysRevB2004} for $\mathrm{UGe_2}$ exhibiting a narrow peak around FM2$\rightarrow$FM1 transition. The latter fact justified our choice of smaller $U / |t| = 3.5$ and $J / |t| = 1.1$.

\bibliography{bibliography}

\end{document}